\documentclass[aps,preprint,nofootinbib,10ppt]{revtex4}
\usepackage{amsfonts}
\usepackage{amsmath}
\usepackage{amssymb}
\usepackage{graphicx,epsfig}
\usepackage{color}
\usepackage[colorlinks=false, pdfstartview=FitV, linkcolor=blue, citecolor=blue, urlcolor=blue]{hyperref}

\begin{document}

\title{Realistic Equations of State for the Primeval Universe}
\author{R. Aldrovandi}
\email{ra@ift.unesp.br}
\affiliation{Instituto de F\'{\i}sica Te\'{o}rica, Universidade Estadual Paulista. Rua
Pamplona 145, CEP 01405-900, S\~{a}o Paulo, SP, Brazil}
\author{R. R. Cuzinatto}
\email{rcuzin@phys.ualberta.ca}
\affiliation{Theoretical Physics Institute, University of Alberta, Edmonton, Alberta,
Canada, T6G 2J1.}
\author{L. G. Medeiros}
\email{leogmedeiros@gmail.com}
\affiliation{Centro Brasileiro de Pesquisas F\'{\i}sicas. Rua Xavier Sigaud 150, CEP
22290-180, Rio de Janeiro, RJ, Brazil}
\pacs{PACS number}

\begin{abstract}
Early universe equations of state including realistic interactions between
constituents are built up. Under certain hypothesis, these equations are
able to generate an inflationary regime prior to the nucleosynthesis period.
The resulting accelerated expansion is intense enough to solve the flatness
and horizon problems. In the cases of curvature parameter $\kappa $ equal to
$0$ or $+1$, the model is able to avoid the initial singularity and offers a
natural explanation for why the universe is in expansion. All the results
are valid only for a matter-antimatter symmetric universe.
\end{abstract}

\volumeyear{2008}
\volumenumber{number}
\issuenumber{number}
\eid{identifier}
\date[Date text]{date}
\received[Received text]{date}
\revised[Revised text]{date}
\accepted[Accepted text]{date}
\published[Published text]{date}
\startpage{1}
\endpage{}
\maketitle



\section{Introduction\label{sec-Intro}}

The gravitational field, as described by General Relativity, couples to all
types of energy: rest masses, kinetic terms and interaction terms.
Relativistic cosmology is, in consequence, deeply concerned with such
sources. The kinetic terms and the rest-masses are currently taken into
account by assuming ideal cosmic fluids constituted by ultrarelativistic
and/or non-relativistic matter. Solutions of this type are found in standard
texts \cite{nar,wein} and -- when multi-component fluids are considered --
in several papers, e.g., \cite{Solutions}. Interaction terms are commonly
used only in perturbative models. In fact, using the Boltzmann equation in a
Friedmann-Robertson-Walker (FRW) background, the inhomogeneities both in the
cosmic microwave background (CMB) and in the matter content \cite{Dod,kolb}
are studied, for comparison with the observational data \cite%
{Wmap,LargeScale}. Notwithstanding, interactions are not considered as
direct sources of gravitation in this line of research \cite{artigo 1}.

Of course, some well-known proposals do consider interaction processes as
direct sources of gravitation. They are usually related to the accelerated
expansion regimes: present-day dynamics \cite{SuNo} or inflation \cite%
{Liddle}. Nevertheless, these theories do not actually consider the
fundamental interactions (electromagnetic, weak and strong) between
particles in the source constituents. Indeed, in the standard inflationary
approaches \cite{Guth,AlbreSten,Linde} an accelerated expansion is obtained
through self-interaction processes of scalar \textit{inflaton} fields $\phi
_{i}\left( x\right) $. This self-interaction is chosen so as to produce just
those features which are necessary to describe the early accelerated regime
\cite{Liddle}. The phenomenological explanations for the present-day
acceleration include, among others, \emph{(i)} the quintessence models \cite%
{CalDaveStein,Stein,Zlatev,FraRos,PeRa}, which roughly follow the same lines
of the inflationary theory; \emph{(ii)} models of matter and dark-energy
unification via Chapligyn-like equations of state (EOS) \cite{Orfeu} or
through equations of the Van der Walls type \cite{Capozzielo,Kremer}; \emph{%
(iii)} models of mass-varying-neutrino type, which couple neutrinos to a
quintessence scalar field \cite{Motta}; \emph{(iv)} models introducing
interactions in the energy conservation equation \cite{Gabi,NelsonPinto}.

A formal procedure has been recently proposed for including the
(fundamental) interactions as direct sources of gravitation in the
cosmological context \cite{artigo 1}. Imported from equilibrium statistical
mechanics, this formalism allows the construction of realistic equations of
state for both relativistic \cite{Dashen,Reichl} and non-relativistic
systems \cite{Pat,Beth}. Our objective here is to find primeval cosmic fluid
EOS taking into account \textit{physically realistic} interaction processes
between the constituent particles, in addition to their kinetic and
rest-mass terms. In particular, we shall examine their effect on the scale
factor evolution. It will be shown that under certain hypothesis and
approximations an early accelerated regime can be obtained as a consequence
of these interacting processes.

The idea of building realistic equations of state considering
interaction between elementary particles in the primeval universe is
not new. Actually, during the late 1970's and the beginning os the
80's a series of papers by Bugrii, Trushevsky and Beletsky
\cite{ucraone,ucratwo,ucrathree,ucrafour} discussed the construction
of the high-energy EOS and their application to
the pre-nucleosynthesis universe.\footnote{%
The authors are thankful to an unknown referee for calling their attention
to these works.} Nevertheless, their treatment is different from the one
developed here as we will make clear some pages ahead.

The paper is organized as follows. Section \ref{sec-Pns} reviews the general
features of the early universe in its standard presentation \cite%
{nar,wein,Dod,kolb,Liddle}. Section \ref{sec-SisInt} presents some results
from equilibrium statistical theory of interacting systems. Specifically,
the coefficients appearing in the perturbative fugacity expansions are
expressed in terms of the scattering matrix operator $\hat{S}$. In addition,
the matrix $S_{2}$ describing the two-particle scattering is associated to
the experimentally observable phase-shifts. The goal of Section \ref%
{sec-EoS_Pns} is to construct realistic EOS\ for the pre-nucleosynthesis
universe and discuss the hypothesis and approximations undertaken. In
Section \ref{sec-Conseq},\ the more direct cosmological consequences coming
from these equations are examined, including the effect of driving an
accelerated expansion (inflationary era). Section \ref{sec-Final} contains
some final comments. Details of a too technical nature, as well as some
data, have been relegated to appendices.


\section{General features of the pre-nucleosynthesis universe \label{sec-Pns}%
}


The period of the Universe evolution going under the name ``Early Universe''
covers many different and physically significant events. Indeed, it is usual
to consider both the matter-radiation decoupling ($kT_{\gamma }\sim 0.1 $ $%
eV $) and the electroweak transition ($kT_{\gamma }\sim 100$ $GeV$)\ as
belonging to that period. This work is concerned with a specific interval
within it, namely the pre-nucleosynthesis period (PNS). We shall define PNS
as the period immediately before the nucleosynthesis of the light elements,
when the nucleons (protons and neutrons) are in thermodynamical equilibrium
with the rest of the cosmic fluid, i.e., $kT_{\gamma }\gtrsim 20$ $MeV$. So,
in principle, any energy value larger than $20$ $MeV$ belongs to PNS;
nevertheless, it will be enough to restrict our working frame to energies of
a few hundreds of $MeV$. In consequence, PNS will in what follows actually
mean the interval $20$ $MeV\leq kT\lesssim 300$ $MeV$.

Present-day universe is dominated by a dark energy component $\Lambda $\
plus a non-relativistic contribution formed by ordinary baryons $b$\ and
dark-matter $dm$; the ultrarelativistic components -- photons $\gamma $\ and
neutrinos $\nu $\ -- are quite negligible. As one goes back in time,
however, this situation changes drastically. Indeed, the evolution equations
for ultrarelativistic and the nonrelativistic matter \cite{nar,wein} in
terms of the expansion parameter,
\begin{equation}
\rho _{\gamma ,\nu }=\rho _{\gamma 0,\nu 0}\left( \frac{a_{0}}{a}\right) ^{4}%
\text{ \ \ \ \ and \ \ \ }\rho _{b,DM}=\rho _{b0,DM0}\left( \frac{a_{0}}{a}%
\right) ^{3},  \label{C1_10}
\end{equation}%
together with the Friedmann equations,%
\begin{equation}
\left( \frac{\dot{a}}{a}\right) ^{2}=\frac{8\pi G}{3}\rho -\frac{\kappa }{%
a^{2}}+\frac{\Lambda }{3}\text{ \ \ \ and \ \ \ }\left( \frac{\ddot{a}}{a}%
\right) =-\frac{4\pi G}{3}\left( \rho +3p\right) +\frac{\Lambda }{3},
\label{C1_11}
\end{equation}%
show that the ultrarelativistic contribution overcomes that of the other
components at energies corresponding to $kT_{\gamma }\sim 1$ $eV$.\footnote{%
We will see in Section \ref{sec-EoS_Pns}\ that this statement must be
qualified when the interaction processes are taken into account.} The
cosmological period dominated by ideal ultrarelativistic particles is
usually called the \textit{radiation era}.

To find out the energetically relevant constituents during the PNS period we
must notice that, as we turn to the past and the energy $kT$\ increases,
pair-production processes become more and more frequent, generating a large
variety of particles. The main scenario is that of ultrarelativistic
particles generating non-relativistic ones. If we restrict ourselves to a
few hundreds of $MeV$, the relevant particles are:

\begin{itemize}
\item \textit{fundamental bosons:} photons $\left( \gamma \right) $;

\item \textit{leptons:\ }electrons $\left( e^{-}\right) $, positrons $\left(
e^{+}\right) $, muons $\left( \mu ^{-}\right) $, antimuons $\left( \mu
^{+}\right) $, electronic and muonic neutrinos $\left( \nu _{e},\nu _{\mu
}\right) $ and electronic and muonic antineutrinos $\left( \bar{\nu}_{e},
\bar{\nu}_{\mu }\right) $;

\item \textit{hadrons:\ }pions $\left( \pi ^{+},\pi ^{-},\pi ^{0}\right) $,
kaons $(K^{+},K^{-},K^{0},\bar{K}^{0})$, nucleons $\left( p,n\right) $ and
antinucleons $\left( \bar{p},\bar{n}\right) $.
\end{itemize}

Concerning this list, it is important to emphasize that: \textit{(i)} the
particles taken into account are those with rest-mass bellow $1$ $GeV$; and,
\textit{(ii)} the hadrons considered are only those which are stable by the
strong interaction. Furthermore, we assume matter-anti-matter symmetry.

Among the particles cited above, those that are ultrarelativistic at $20$ $%
MeV$\ are $\gamma $, $e^{\pm }$, $\nu _{e}$, $\nu _{\mu }$, $\bar{\nu}_{e}$
and $\bar{\nu}_{\mu }$; the remaining ones are non-relativistic. By
arguments given in classical texts \cite{nar,wein,kolb}, the chemical
potentials involved in the PNS reactions are zero, i.e., $\mu _{i}\simeq 0$
for every species ($i=\gamma $, $e^{\pm }$, $\mu ^{\pm }$, $\pi ^{\pm }$, $%
K^{\pm }$, ...).

Thermal equilibrium is warranted as long as every reaction rate $\Gamma
_{i}(kT)$ of each given component $i$ with all the other particles is much
larger than the universe expansion rate. This last, in turn, is measured by
the Hubble function $H(kT)$. Therefore, when the condition%
\begin{equation}
\Gamma (T)\gg H(T)  \label{C1_22}
\end{equation}%
is satisfied, thermodynamical \textit{quasi-static expansion} holds. In
fact, this may be verified for each variety by remembering that $\Gamma
=n\sigma \left\vert \vec{v}\right\vert $, where $n$ is the target-particle
density and $\sigma \left\vert \vec{v}\right\vert $\ is the interaction
cross-section times the relative velocity of the particles. The presence of
thermal and chemical equilibria in PNS is extremely convenient, as it
enables us to adopt, in the comoving reference frame, the usual statistical
mechanics on the $\mathbf{E}^{3}$ manifold (the phase space measure is given
simply by $\frac{1}{\left( 2\pi \right) ^{3}}\iint d^{3}xd^{3}p$).

The energy density turning up in (\ref{C1_11}) is \cite{nar,kolb}
\begin{equation}
\rho _{UR}=\rho _{\gamma }+\rho _{e^{-}}+\rho _{e^{+}}+\rho _{\nu _{e}}+\rho
_{\nu _{\mu }}+\rho _{\bar{\nu}_{e}}+\rho _{\bar{\nu}_{\mu }}=\frac{9}{2}\,
\rho _{\gamma }=\frac{9\pi ^{2}}{30}(kT)^{4}  \label{C1_17}
\end{equation}%
and can be used to obtain the Hubble function as a function of $kT$:%
\begin{equation}
H(kT)=\left( \frac{\dot{a}}{a}\right) =\sqrt{\frac{4\pi ^{3}}{5}} \, \frac{%
(kT)^{2}}{m_{Pl}},  \label{C1_18}
\end{equation}%
where $m_{Pl}\equiv G^{-1/2}=1.221\times 10^{22}$ $MeV$. The relation
between the scale factor and the energy $kT$ is determined by comparing (\ref%
{C1_17})\ to (\ref{C1_10}):%
\begin{equation}
kT=\frac{A}{a} \,\, .  \label{C1_19}
\end{equation}%
$A$ is a constant which depends on the present--day values $T_{0}$ and $%
a_{0} $. Substituting this last equation into (\ref{C1_18}) and solving the
resulting differential equation -- imposing the initial condition $a(t=0)=0$%
\ -- leads to%
\begin{equation}
a(t)=A\left( \frac{16\pi ^{3}}{5}\right) ^{1/4}\left( \frac{t}{m_{Pl}}%
\right) ^{1/2}.  \label{C1_20}
\end{equation}%
From (\ref{C1_19}) follows then the relation between the energy $kT$ and the
cosmological time $t$:
\begin{equation}
kT=\left( \frac{16\pi ^{3}}{5}\right) ^{-1/4}\left( \frac{t}{m_{Pl}}\right)
^{-1/2}  \label{C1_21}
\end{equation}%
The last four equations determine the evolution of the primeval universe in
a radiation--dominated era.


\section{Statistics of interacting systems\label{sec-SisInt}}


It was discussed elsewhere \cite{artigo 1} how to construct equations of
state for an interacting system of particles within the standard ensemble
formalism of statistical mechanics, and how this could be applied to a
simple model relevant to cosmology. In what follows, we will present a brief
review of the information needed here.\ We also indicate Refs. \cite{Dashen}%
, \cite{Beth}, \cite{mayer} and \cite{LeeYang} for further information.

The grand canonical partition function $\Theta $\ is expressed as
\begin{equation}
\Theta (z,V,T)=\sum\limits_{N=0}^{\infty }Q_{N}(V,T)z^{N},  \label{s1}
\end{equation}%
where $V$ is the volume, $T$ is the temperature, $z=e^{\mu /kT}$ is the
fugacity and $Q_{N}(V,T)$ is the N-particle canonical partition function.
The chemical potential $\mu =\mu ^{NR}+m$ is composed by the rest mass $m$
plus the non-relativistic chemical potential $\mu ^{NR}$. This last is that
usually found in standard statistical mechanics texts, e.g. \cite{Pat}.
Here, we are concerned only with one-component systems. For the cases of
more constituents, see Appendix \ref{ap-Multicomp}.

The thermodynamical quantities -- pressure $p$, energy density $\rho $,
numerical density $n$, etc. -- are related to the grand canonical potential
\begin{equation}
\Omega (z,T)\equiv \frac{1}{V}\ln \Theta (z,V,T)=\sum\limits_{N=1}^{\infty
}b_{N}z^{N},  \label{s2}
\end{equation}%
written in terms of the \textit{cluster integrals} $b_{N}$. The
thermodynamical limit has already been taken in Eq. (\ref{s2}), so that the $%
b_{N}$ are functions of the temperature solely (the $V$ dependence
disappears). According to Dashen, Ma and Bernstein \cite{Dashen}, the
cluster integrals are calculated from the $S$-matrix as follows:
\begin{equation}
b_{N}-b_{N}^{(0)}=\frac{g_{N}}{V}\int \frac{e^{-\beta E}}{4\pi i}Tr\left(
\hat{A}\hat{S}^{-1}\frac{\overleftrightarrow{\partial }}{\partial E}\hat{S}%
\right) _{\tilde{c}_{N}}dE,  \label{s3}
\end{equation}%
where $\beta =1/kT$, $g_{N}$ counts the degeneracy coming from internal
degrees of freedom, $\hat{A}$ symmetrizes (antisymmetrizes) the bosonic
(fermionic) states, $\hat{S}$ is the scattering matrix operator and $%
b_{N}^{(0)}$ is the cluster integral of the non-interacting quantum system.
The subscript $\tilde{c}_{N}$ represents all the $N$-particle connected
diagrams\footnote{%
An $N$-particle connected diagram is a graphic representation of $N$ balls
linked directly or indirectly by lines which represent the correlations
coming from interactions or statistical effects.} in which the interaction
occurs at least once. In (\ref{s3}), it was used the short-cut
\begin{equation}
\hat{S}^{-1}\frac{\overleftrightarrow{\partial }}{\partial E}\hat{S}\equiv
\hat{S}^{-1}\frac{\partial \hat{S}}{\partial E}-\frac{\partial \hat{S}^{-1}}{%
\partial E}\hat{S}.  \label{s4}
\end{equation}%
By using (\ref{s2}) $p$, $n$ and $\rho $ are determined as series in the
fugacity:
\begin{equation}
\frac{p(z,T)}{kT}=\sum\limits_{N=1}^{\infty }b_{N}z^{N}~;\text{ \ }%
n(z,T)=\sum\limits_{N=1}^{\infty }Nb_{N}z^{N};\text{ \ }\rho
(z,T)=-\sum\limits_{N=1}^{\infty }\frac{\partial b_{N}}{\partial \beta }%
z^{N}.  \label{D32A}
\end{equation}%
This is the parametric form of the equations of state.

An alternative description is given by the pressure $p(n,T)$ and the energy
density $\rho (n,T)$ written in terms of $n\left( z,T\right) $. They are
obtained by inversion of series $n\left( z,T\right) $ and substitution into $%
p(z,T)$ and $\rho (z,T)$. This results in the \textit{virial expansion}:
\begin{equation}
\frac{p(n,T)}{kT}=\sum_{l=1}^{\infty }a_{l}(T)n^{l};\ \ \ \frac{\rho (n,T)}{%
\left( kT\right) ^{2}}=\sum_{l=1}^{\infty }c_{l}(T)n^{l},  \label{D33}
\end{equation}%
with $a_{l}$ and $c_{l}$ representing the virial coefficients for the
pressure and the energy density. These coefficients are completely
determined by the $b_{N}$. Appendix \ref{ap-Multicomp} presents the explicit
forms of $a_{l}$ and $c_{l}$ for a two-component system.

In practice, it is not possible to really sum any of the series (\ref{D32A}-%
\ref{D33}). Therefore, the choice of using $p$ and $\rho $ in terms of $z$
or $n$ depends on the perturbative characteristics of each series and on the
system under study. We will return to this question in Section \ref%
{sec-EoS_Pns}.


\subsection{The second coefficient of the fugacity expansion}


The elastic interaction between two particles is decomposed into
rotation--invariant sectors, so that each part depends only on their
relative distance $r$ (central interaction):
\begin{equation}
\hat{V}^{a}(r)\equiv \hat{H}^{a}-\hat{H}^{(0)a},  \label{Rcinquenta
oito}
\end{equation}
index $a$ standing for the other types of invariance: spin, isospin, charge
conjugation, etc.\footnote{%
\, For instance, proton-neutron interaction ($pn$) related to total spin $S$%
, total angular momentum $J$ and orbital angular momentum $L$ can be
represented by a central interaction operator such as $\hat{V}%
_{^{2S+1}L_{J}}^{pn}(r)$.} The scattering matrix $S_{2}^{a}$\ depends then
only on the energy and the angle between the initial and final momenta. In
this angular momentum representation, the $\hat{S}_{2}^{a}$\ operator can be
written solely in terms of the phase shifts $\delta _{l}^{a}(k)$:
\begin{equation}
\left\langle k^{\prime }l^{\prime }m^{\prime }\right\vert \hat{S}%
_{2}^{a}\left\vert klm\right\rangle =e^{2i\delta _{l}^{a}(k)}\delta
_{k^{\prime },k}\delta _{l^{\prime },l}\delta _{m^{\prime },m}.
\label{Rsetenta quatro}
\end{equation}

The symmetrization performed by operator $\hat{A}$\ in (\ref{s3}) must
account for the complete state associated with $a$. It is possible to
transfer this symmetrization instruction to the index $a$. Once this is
done, the cluster integral of the two-particle system is written as%
\begin{equation}
b_{2}-b_{2}^{(0)}=\sum\limits_{a}\frac{g_{2}^{a}}{V}\int \frac{dE}{4\pi i}%
e^{-\beta E}Tr\left[ \left( \hat{S}_{2}^{a}\right) ^{-1}\frac{%
\overleftrightarrow{\partial }}{\partial E}\hat{S}_{2}^{a}\right] ,
\label{Rsetenta oito}
\end{equation}%
where $g_{2}^{a}$ counts the degeneracy degree of the duly symmetrized
states.

If we carry out a coordinate transformation to the center of mass, the state
will behave just like a free state (plane wave), that is, it will obey the
free Hamiltonian. In order to explore this fact, we remember some properties
of the relativistic invariant $s$:%
\begin{align}
\omega ^{2}& \equiv s=\left( p_{1\mu }+p_{2\mu }\right)
^{2}=m_{1}^{2}+m_{2}^{2}+2p_{1\mu }p_{2}^{\mu }  \notag \\
& =E_{1}^{2}+E_{2}^{2}+2E_{1}E_{2}-\left( \vec{p}_{1}+\vec{p}_{2}\right)
^{2}=E^{2}-\vec{P}^{2},  \label{Rsetenta nove}
\end{align}%
$m_{1}$ and $m_{2}$ are the masses of the two particles,\ $E=E_{1}+E_{2}$ is
the sum of their energy, and the momentum of the center of mass is $\vec{P}%
\equiv \vec{p}_{1}+\vec{p}_{2}$. Since $E$ and $\vec{P}$ are the energy and
the momentum of the two-particle cluster, $\omega $ may be understood as the
\textit{mass} of this cluster and encapsulates all the information about the
interaction. As the interaction calculated by (\ref{Rsetenta oito}) does not
depend on the coordinates of the center of mass, the integration and the
differential operator with respect to $E$ are reexpressed as%
\begin{equation}
b_{2}-b_{2}^{(0)}=\sum\limits_{a}\frac{g_{2}^{a}}{V}\int d^{3}R\int \frac{%
d^{3}P}{\left( 2\pi \right) ^{3}}\int \frac{d\omega }{4\pi i}e^{-\beta \sqrt{%
\vec{P}^{2}+\omega ^{2}}}Tr\left[ \left( \hat{S}_{2}^{a}\right) ^{-1}\frac{%
\overleftrightarrow{\partial }}{\partial \omega }\hat{S}_{2}^{a}\right] .
\label{Roitenta quatro}
\end{equation}%
The next step is to integrate on $d^{3}P$:
\begin{equation}
b_{2}-b_{2}^{(0)}=\sum\limits_{a}\frac{g_{2}^{a}}{2\beta \pi ^{2}}%
\int\limits_{M}^{\infty }\omega ^{2}K_{2}(\beta \omega )\frac{1}{4\pi i}Tr%
\left[ \left( \hat{S}_{2}^{a}\right) ^{-1}\frac{\overleftrightarrow{\partial
}}{\partial \omega }\hat{S}_{2}^{a}\right] d\omega ,  \label{s5}
\end{equation}%
where $M=m_{1}+m_{2}$ is the minimum value assumed by $\omega $; $%
K_{2}(\beta \omega )$ is the modified Bessel function. Using (\ref{Rsetenta
quatro}), the trace in the angular momentum representation is

\begin{equation}
Tr\left[ \left( \hat{S}_{2}^{a}\right) ^{-1}\frac{\overleftrightarrow{%
\partial }}{\partial \omega }\hat{S}_{2}^{a}\right] =4i\sum\limits_{l=0}^{%
\infty }\left( 2l+1\right) \frac{\partial \delta _{l}^{a}(\omega )}{\partial
\omega } \,\, .  \label{Rnoventa um}
\end{equation}%
Equation (\ref{s5}) becomes:%
\begin{equation}
b_{2}-b_{2}^{(0)}=\frac{1}{2\beta \pi ^{3}}\sum\limits_{a}\sum%
\limits_{l=0}^{\infty }g_{2}^{a}\left( 2l+1\right) \int\limits_{M}^{\infty
}\omega ^{2}K_{2}(\beta \omega )\frac{\partial \delta _{l}^{a}(\omega )}{%
\partial \omega }d\omega .  \label{Rnoventa dois A}
\end{equation}%
This equation, which reduces to the Beth~--~Uhlenbeck \cite{Beth,Pat} case
in the non-relativistic limit, is the fundamental tool to obtain the EOS for
the pre-nucleosynthesis universe.


\section{Equations of state for the pre-nucleosynthesis universe \label%
{sec-EoS_Pns}}


The EOS to be built here are supposedly realistic because they account for
the interactions. Nevertheless, we will not actually consider all the four
fundamental interactions.

The gravitational interaction is accounted for only through Einstein (more
specifically, Friedmann) equations. We will neglect any possible change in
the description of statistical mechanics due to the curved background of the
cosmic manifold.\footnote{%
\ Ref. \cite{AlBelPe}\ discusses this subject for the de Sitter solution.}

The electromagnetic interaction, though responsible for the thermalization
of the charged particles, will not be relevant to the primeval EOS, since:
(i) the shielding effect due to the existence of opposite charges makes it
possible to consider the effective interaction as of short range, and the
fluid as neutral; (ii) the coupling constant of the electromagnetic
interaction is relatively small ($e^{2}\sim 1/137$). The mean interaction
energy between charged particles is nearly two orders of magnitude lesser
than their mean kinetic energies \cite{nar}. The weak interaction, though
responsible for the thermalization of the neutral particles, is of
short-range, and its intensity is much smaller \cite{Grif} than the strong
interaction. It will be neglected. Due to its high coupling constant, and
despite its short-range, the strong (``hadronic'') interaction is dominant
in the PNS period. As a matter of fact, it is the only one that will be
taken into account (see Appendix \ref{ap-Espalha}).

Justified by these considerations, we divide the relevant particles for the
PNS in three categories:

\begin{enumerate}
\item Ideal ultrarelativistic particles: $\gamma ,e^{-},e^{+},\nu _{e},\nu
_{\mu },\bar{\nu}_{e},\bar{\nu}_{\mu }$;

\item Ideal relativistic particles: $\mu ^{-},\mu ^{+}$;

\item Interacting relativistic particles: $\pi ^{+},\pi ^{-},\pi
^{0},K^{+},K^{-},K^{0},\bar{K}^{0},p,\bar{p},n,\bar{n}$.
\end{enumerate}

The equations of state for such a system of particles is the sum of the
contributions of each species to quantities $p$ and $\rho $, keeping in mind
what has been said in Section \ref{sec-SisInt} whenever interactions are
important. Both $p$ and $\rho $\ can be written as functions of $(z,T)$ or
in terms of $(n,T)$ -- Eqs. (\ref{D32A}-\ref{D33}). Thus, there are four
pairs $\left\{ p,\rho \right\} $\ given by the combination of $p(z,T)$, $%
p(n,T)$, $\rho (z,T)$\ and $\rho (n,T)$. We have to decide which combination
is the most suitable.

There are strong theoretical arguments (based on the series convergences)
and outstanding experimental indications \cite{artigo 1,Hirshefeld} that the
virial expansion $p(n,T)$\ is the most convenient form for the pressure. The
choice between $\rho (z,T)$ and $\rho (n,T)$ is more subtle. We shall keep
the form $\rho (z,T)$ since it maintains the probabilistic notion inherited
by statistical mechanics. In fact, the energy density is a mean weighted by
the Boltzmann factor, and in the case of the grand canonical ensemble it
reads%
\begin{equation}
\rho =\frac{E}{V}=\frac{1}{\Theta (V,T,z)}\sum\limits_{a,r}\frac{E_{a}}{V}%
e^{-\beta E_{a}}z^{N_{r}}.  \label{s7}
\end{equation}%
Comparing this equation to (\ref{s2}) and (\ref{D32A}) one sees that only $%
\rho (z,T)$\ preserves the probabilistic character order by order.

Hence, the EOS for the relativistic particles (interacting or not) will be
formed by the virial series for the pressure and the fugacity series for the
energy density: $\left\{ p(n,T),\rho (z,T)\right\} $. Only the first terms
of the expansions will be considered, as only two-by-two interactions can be
calculated or experimentally measured. Following the classification above,
the pressure and energy density will be
\begin{eqnarray}
p &=&p_{Had}+p_{R}+p_{UR}=p_{Had}+p_{R}+\frac{\rho _{UR}}{3},  \label{C5_3}
\\
\rho &=&\rho _{Had}+\rho _{R}+\rho _{UR},  \label{C5_4}
\end{eqnarray}%
where the labels $Had$, $R$ and $UR$ represents the hadrons, the ideal
relativistic particles and the ideal ultrarelativistic particles.


\subsection{EOS for the hadrons \label{sec-EoS_Had}}

Pions $\pi ^{+},\pi ^{-},\pi ^{0}$, kaons $K^{+},K^{-},K^{0},\bar{K}^{0}$\
and nucleons $N = p,\bar{p},n,\bar{n}$\ exhibit spin, isospin and charge
conjugation symmetries under strong interactions (Appendix \ref{ap-Espalha}%
). For this reason, the set of hadrons can be treated as an interacting
system of three components -- $\pi $, $K$\ and $N$\ -- whose EOS are derived
from (\ref{D41c}) and (\ref{D42a}). The energy density will be
\begin{eqnarray}
\rho _{Had}(z_{\pi },z_{K},z_{N},kT) &=&\left( kT\right) ^{2}\left[ \dot{b}%
_{1\pi }z_{\pi }+\dot{b}_{1K}z_{K}+\dot{b}_{1N}z_{N}+\right.  \notag \\
&&\left. +\dot{b}_{2\pi \pi }z_{\pi }^{2}+\dot{b}_{2KK}z_{K}^{2}+\dot{b}%
_{2NN}z_{N}^{2}+\right.  \notag \\
&&\left. +\dot{b}_{2\pi K}z_{\pi }z_{K}+\dot{b}_{2\pi N}z_{\pi }z_{N}+\dot{b}%
_{2KN}z_{K}z_{N}\right] ,  \label{C5_6}
\end{eqnarray}%
where the dot indicates differentiation with respect to $kT$. To this point
we have not used the fact (already discussed) that $\mu _{\pi }\simeq \mu
_{K}\simeq \mu _{N}\simeq 0$ during the pre-nucleosynthesis period, and
consequently $z_{\pi }\simeq z_{K}\simeq z_{N}\simeq 1$. The pressure is
\begin{eqnarray}
p_{Had}(n_{\pi },n_{K},n_{N},kT) &=&kT\left( a_{1\pi }n_{\pi
}+a_{1K}n_{K}+a_{1N}n_{N}+\right.  \notag \\
&&\left. +a_{2\pi \pi }n_{\pi }^{2}+a_{2KK}n_{K}^{2}+a_{2NN}n_{N}^{2}+\right.
\notag \\
&&\left. +a_{2\pi K}n_{\pi }n_{K}+a_{2\pi N}n_{\pi
}n_{N}+a_{2KN}n_{K}n_{N}\right) ,  \label{C5_8}
\end{eqnarray}%
with the numerical densities given by
\begin{eqnarray}
n_{\pi }(z_{\pi },z_{K},z_{N},kT) &=&b_{1\pi }z_{\pi }+2b_{2\pi \pi }z_{\pi
}^{2}+b_{2\pi K}z_{\pi }z_{K}+b_{2\pi N}z_{\pi }z_{N},  \label{C5_9a} \\
n_{K}(z_{\pi },z_{K},z_{N},kT) &=&b_{1K}z_{K}+2b_{2KK}z_{K}^{2}+b_{2\pi
K}z_{\pi }z_{K}+b_{2KN}z_{K}z_{N},  \label{C5_9b} \\
n_{N}(z_{\pi },z_{K},z_{N},kT) &=&b_{1N}z_{N}+2b_{2NN}z_{N}^{2}+b_{2\pi
N}z_{\pi }z_{N}+b_{2KN}z_{K}z_{N}.  \label{C5_9c}
\end{eqnarray}%
The explicit form of the ideal and interaction-related terms are given in
Appendix \ref{ap-Multicomp}. The rest mass adopted are: $m_{\pi }=0.1396$ $%
GeV$, $m_{K}=0.4957$ $GeV$ and $m_{N}=0.93826$ $GeV$.

In principle, the term $b_{2}$\ should contain all the interaction processes
involving the two particles to which it refers. That is, it should describe
the elastic and inelastic scatterings as well as the bound states. Actually,
it is possible to argue that, in the construction of the EOS for the
pre-nucleosynthesis period, bound states and inelastic processes are
negligible when compared to the elastic channels.

Taking into account just the elastic processes, the six interaction terms ($%
\pi \pi $, $\pi K$, $\pi N$, $KK$, $KN$ and $NN$) are given by equations of
type (\ref{Rnoventa dois A}). The explicit expression of the cluster
integrals are obtained with the help of all the phase-shift data sets -- see
Appendix \ref{ap-Espalha} for an example. It is necessary to perform an
integration involving the modified Bessel function $K_{2}$ and derivatives
of the phase-shifts. The six integrations have been done numerically with
the software \emph{Mathematica 5.0}. Before that, we carefully put back the
constants $c$ and $\hbar $ in such a way that the length and energy units
were $fm$ and $GeV$, respectively.

The higher limits of the integrals are different for each $b_{2}$, since the
phase-shift data sets are obtained in different energy intervals. Still,
most of them lie in the interval from $1$ $GeV$ to $2.5$ $GeV$. As the
relevant temperature values for our studies are in the range $20$ $MeV\leq
kT\lesssim 300$ $MeV$, the chosen values for the\ higher limits are good
enough to include all the main contribution from the integral kernels.
Besides, the Bessel function $K_{2}$ assures the fast decrease of the kernel
values. We consider only angular momenta $l=0,1$ and $2$, i.e., the $S$ , $P$
and $D$ contributions (Appendix \ref{ap-Espalha}). Once the coefficients $%
b_{1}$ and $b_{2}$ have been calculated, it is straightforward to obtain the
numerical densities (\ref{C5_9a}), (\ref{C5_9b}) and (\ref{C5_9c}), and the
equations of state for the hadrons, Eqs. (\ref{C5_6}) and (\ref{C5_8}).

The approach described above is not the only one for building hadronic EOS
at few hundreds of $MeV$. There is also the treatment called Hadron
Resonance Gas (HRG) \cite{Munzinger,Andro,tawfik2,Cheng} describing the
Fireballs created in the Heavy Ion Colliders (SPS -- Super Proton Synchroton
-- and RHIC -- Relativistic Heavy Ion Collider). This technique agrees
qualitatively with the Lattice QCD calculations \cite{Cheng} and it
reproduces the abundance of the observed particles relatively well \cite%
{Andro}. The HRG modeling is equivalent to ours as long as the ideal terms
are concerned, but the procedure of including the interaction is rather
different. The Hadron Resonance Gas accounts for interaction by introducing
a excluded volume term \emph{a la} Van der Walls \cite{Yen} and by
considering the resonance contributions \cite{Munzinger,Andro}.\footnote{%
The HRG model is similar to the Hagedorn's statistical bootstrap model \cite%
{Hag} in the aspect that both cases include the hadronic interactions
primary through the resonances.} On the other hand, our work accounts for
interactions through the phase shifts. Because of this, other effects --
besides the ones related to resonances and hard cores (excluded volumes) --
are considered. Reference \cite{Delta} analyzes these features in detail for
the case of $\Delta $ resonance in the $\pi N$ interaction. It is worth to
emphasize that in our approach the resonances are only characteristics of
the elastic scattering between stable hadrons; the resonances are not taken
as particles.

In connection with what we said in the Introduction: the construction of
hadronic EOS through the S-Matrix was already implemented in Refs. \cite%
{ucraone,ucratwo}. However, the only processes considered there (e.g. \cite%
{ucraone}) are those including the nucleons (scattering $NN$). The present
work is more constructive and complete in this sense: all the important
statistical quantities -- pressure, energy density and numerical density --
are obtained from the partition function which accounts for all the relevant
processes of interaction two by two (scattering $\pi \pi $, $\pi K$, $\pi N$%
, $KK$, $KN$ and $NN$).


\subsection{EOS for the ideal particles}

The ideal (non-interacting) particles are divided into two categories:
ultrarelativistic ($\gamma , e^{-},\allowbreak e^{+}, \nu _{e}, \nu _{\mu },
\bar{\nu}_{e}, \bar{\nu}_{\mu }$) and relativistic ($\mu ^{-},\mu ^{+}$).
The ultrarelativistic sector is relatively simple, and has been discussed in
Section \ref{sec-Pns}. The calculation for the relativistic sector is
analogous to what we have done for the hadrons. The energy density, for
instance, is
\begin{equation}
\rho _{R}(z_{\mu },kT)=\left( kT\right) ^{2}\left[ \dot{b}_{1\mu
}^{(0)}z_{\mu }+\dot{b}_{2\mu }^{(0)}z_{\mu }^{2}\right] .  \label{C5_28}
\end{equation}%
The $b_{1\mu }^{(0)}$ and $b_{2\mu }^{(0)}$ are the ideal terms determined
by (\ref{Aj11}) and (\ref{Aj12}), with $m_{\mu }=0.10566$ $GeV$. As in the
hadronic case, we set $z_{\mu }\simeq 1$ during the PNS. The pressure is%
\begin{equation}
p_{R}(n_{\mu },kT)=kT\left[ a_{1\mu }n_{\mu }+a_{2\mu }n_{\mu }^{2}\right]
\text{ \ \ \ \ with \ \ \ } n_{\mu }(z_{\mu },kT)=b_{1\mu }^{(0)}z_{\mu
}+2b_{2\mu }^{(0)}z_{\mu }^{2} \, .  \label{C5_30}
\end{equation}


\subsection{The complete EOS}

\begin{figure}[ht]
\begin{center}
\includegraphics[height=6.4cm, width=8.5cm]{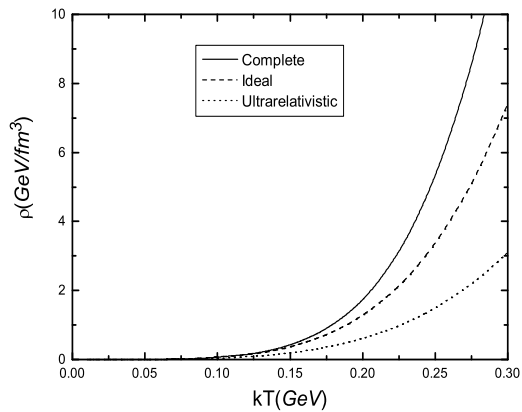}
\end{center}
\caption{Plot of the energy density $\protect\rho $ in $GeV/fm^{3}$ as a
function of the thermal energy $kT$ given in $GeV$. Three curves are
presented: the complete energy density which includes the effects of all the
relevant particles and their interactions (full line); energy density of all
particles (ultrarelativistic, relativistic and hadronic) without considering
interactions, i.e., all the particles are taken as ideal ones (dashed line);
energy density for the ultrarelativistic particles (dotted line).}
\label{fig1}
\end{figure}

The complete EOS for the energy density $\rho $ is obtained by substituting
Eqs.(\ref{C5_6}), (\ref{C1_17}) and (\ref{C5_28}) into (\ref{C5_4}). The
plot of $\rho(kT)$ is shown in Figure \ref{fig1}. The three curves exhibit
similar features: all are positive and increase monotonically. The
ultrarelativistic components dominate up to $\approx 0.12$ $GeV$. From $0.12$
$GeV$ to $0.16$ $GeV$, the other particles (mainly $\pi $ and $\mu $) become
important. At $0.275$ $GeV$ the ultrarelativistic, ideal-relativistic, and
interacting sectors correspond respectively to $25\%$, $33\%$ and $42\%$ of
the total energy density. 

\bigskip

The complete EOS for the pressure $p$ comes from the substitution of Eqs.( %
\ref{C5_8}), (\ref{C1_17}) and (\ref{C5_30}) into (\ref{C5_3}). Figure \ref%
{fig2} shows the plot of function $p=p(kT)$.%

\begin{figure}[ht]
\begin{center}
\includegraphics[height=6.4cm, width=8.5cm]{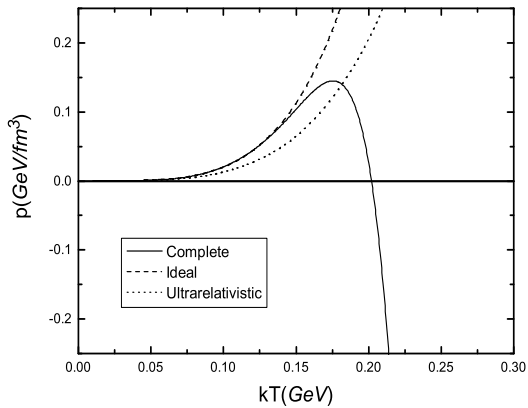}
\end{center}
\caption{Plot of the pressure $p$ in $GeV/fm^{3}$ as a function of the
energy $kT$ given in $GeV$. Interactions are accounted for only in the
complete case (full line). The conclusion is: interaction drags the pressure
to negative values, pointing to cosmic acceleration.}
\label{fig2}
\end{figure}

\bigskip

It can be seen in Figure \ref{fig2} that, up to $0.14$ $GeV$, the three
curves for the pressure present similar behavior: as in the energy density
case, all are positive--valued and increase monotonically. From this point
on the interaction contributions grow, and there is a sudden change in the
slope for the complete system. The curve mitigates its increase rate,
reaches a maximum at $0.175$ $GeV$, and then starts a steep fall. It becomes
negative at around $0.2$ $GeV$ and continues to decrease at a high rate. A
comparative graphic analysis of the effects on $\rho $ and $p$ coming from
the different kinds of interactions and scatterings reveals that the
relevant processes for the determination of $b_{2}$ in scales of $kT\lesssim
300$ $MeV$ are the scatterings $\pi \pi $, $\pi K$ and $\pi N$ related to
the $S$ and $P$ partial waves. Therefore, it is fair enough to take into
account only $l=0,1$ and $2$ when calculating $b_{2}$, just as we did.

A last point we would like to discuss is the validity range of the
expansions $\rho (kT)$ and $p(kT)$. This validity is threatened by three
factors: \textbf{(I)} an eventual phase transition due to instabilities in
the function $p(kT)$; \textbf{(II)} the non-convergence of the pressure and
energy density series, which would not allow the perturbative approach
adopted here; \textbf{(III)} the occurrence of the deconfinement of hadrons
in quarks and gluons, which would change the nature of the particles. Let us
comment on these issues:

\textbf{(I)} There are two necessary and sufficient conditions that assure
the stability of the usual thermodynamical systems \cite{Callen}, namely:
\begin{equation}
\frac{1}{T}\left( \frac{\partial \rho }{\partial T}\right) _{n_{i}}\geq 0%
\text{ \ \ \ and \ \ \ }n_{i}\left( \frac{\partial p}{\partial n_{i}}\right)
_{T}\geq 0 \, ,  \label{s8}
\end{equation}%
with $i=\pi $, $K$, $N$ or $\mu $.

The function $\rho $\ depends only on $kT$\ ($z_{i}=1$ for every species $i$%
), while $p$\ depends on $kT$\ and also on the densities $n_{i}$\ of the
four types of particles. Thus, the second condition in (\ref{s8})\ splits
into four others, all required to guarantee the system stability. From
Figure \ref{fig1}, it is easily seen that the curve $\rho (kT,\{z_{i}\})$
for the complete set always satisfies (\ref{s8}).\footnote{$\left\{
z_{i}\right\} $ represents the set of the four fugacities, all of which
equals one.}

Notwithstanding, the same does not occur with $p\left( kT,\{n_{i}\}\right) $%
. Indeed, as the numerical densities are positive and increasing functions
of $kT$, Figure \ref{fig2} makes clear that the pressure does not obey\
condition (\ref{s8}); i.e., for some energy value higher than $0.14$ $GeV$
the EOS for the pressure becomes unstable. In principle, this instability is
a strong indication of a phase transition and points to a breakdown of the
EOS validity. However, this argument is based on the usual situation of a
thermodynamical system: a gas confined in a recipient of finite volume. By
its very nature, the universe cannot be assumed to have the same
characteristics of such a simple and controlled environment. First: the
universe has no frontiers. In this case, the surface pressure (which is
different from the internal pressure) does not exists. And in many cases are
precisely the surface effects which generate the phase transitions in an
ordinary thermodynamical system. Second: in the context of cosmology,
\textit{the (internal) pressure is a direct source of the gravitational field%
} \cite{artigo 1}. This particular feature produces rather different effects
from those engendered by the pressure in usual thermal systems, for which
one would expect the reduction in the volume as the internal pressure
increases. This notion, spelled by the second relation in (\ref{s8}), is of
great importance, since it prevents the ordinary system from disappearing:
if $\partial p/\partial V$\ were positive, nothing would avoid the collapse
of the gas. On the other hand, in the context of cosmology, an increase of
pressure increases the gravitational attractive effect, and in this sense, a
cosmological system with $\partial p/\partial V<0$ is unstable, i.e., if the
universe is not expanding it tends to collapse. We want to emphasize the
distinction between the effects of the internal pressure in the two types of
systems: in ordinary equilibrium thermodynamics, an increase of $p$ leads to
an increase of the volume $V$ of the recipient containing the gas because
the shocks of the particles against the walls are more and more frequent;
but in cosmology the increase of $p$ strengthens the gravitational field and
produces a net tendency to the reduction of the volume. Hence, it is
reasonable to say that the stability criterion (\ref{s8}) cannot be directly
applied in cosmology, and cannot be used to rule out the above EOS. From the
microscopic point of view, the energy density fluctuations $\delta \rho$
grow fast in regions where $\partial p/\partial \rho \equiv c_{s}^2<0$ ($%
c_{s}$ is named the sound velocity in the media) and the system becomes
unstable. For the usual thermodynamical systems this strongly suggests phase
transition. Our system, the standard cosmological model, is not of this
type, though. The universe expands in a rate determined by the relation
between $p$ and $\rho$. This complicates the analyzes and we can not affirm
that $c_{s}^2<0$ necessarily leads to a phase transition of the system. The
correct treatment of the evolution of $\delta \rho$ should be done along
with the perturbations in the space-time geometry -- FRW metrics. We leave
these investigations for the future.

\textbf{(II)} The convergence of the expansions for $p$ and $\rho $ is
related with the formal aspect of validity of these series and the
correction of the perturbative treatment. This issue is far from trivial,
since is not possible to sum all the terms for any realistic system in a
pure thermodynamical context, let alone in the cosmological framework. The
virial coefficients $a$, or the cluster integrals $b$, depend on the sum of
the connected diagrams representing mutual interactions. And it is extremely
difficult to foresee the behavior of the result when $N$ particles interact.
So, the answer to the question concerning the convergence of $p$ and $\rho $
is inaccessible. Nevertheless, the authors proposed elsewhere \cite{artigo 1}
a toy-model for the interactions in the early universe, computed the
perturbed EOS (until third order) and showed that there is a good indication
of convergence for the pressure series. So, we will assume in this work the
validity of truncating the equations for $p$, $\rho$ and $n$ in the second
order terms.

\textbf{(III)} The QCD coupling constant diminishes as the energy $kT$\
increases. So, one could expect that at a certain critical temperature $%
kT_{c}$, its value becomes sufficiently low to allow for the deconfinement
of the hadronic matter, giving rise to a system composed by quarks and
gluons. If such a transition takes place during the PNS, our model ceases to
be valid beyond $kT_{c}$: the EOS have been calculated assuming that the
fundamental particles are hadrons, not quarks and gluons. According to the
lattice QCD calculations \cite{Lattice}, the critical temperature is
situated between $150$ and $180$ $MeV$. From an experimental perspective,
recent results from the RHIC \cite{RHIC} indicate that a sort of transition
occurs at these temperature, but it is not the deconfinement in a
quark-gluon plasma \cite{QGP}. The experiments run in RHIC found what seems
to be a new state for the nuclear matter. This state would correspond to a
perfect fluid (without viscosity) identified as a CGC (\textit{Color Glass
Condensate}) \cite{CGC}. Unlike the quark-gluon plasma, the CGC has non
negligible correlations, i.e., the nuclear interaction processes are
relevant.

There is no doubt left by the experiments that the hadronic matter in the
heavy ion colliders undergoes a phase transition in energies around $170$ $%
MeV$. But it is possible that this phenomenon does not occur in the
primordial universe (or that it occurs in different energy scales), due to
the various differences between the laboratory and the early universe.
Specifically, the temporal scale of the events in the possible cosmological
QCD transition ($10^{-5}s$) is quite different from the scale of the
transition in the accelerators ($10^{-23}s$) \cite{Vega}. Other differences
-- previously cited -- are the non-existence of border in the cosmological
system and the role of pressure as direct source of gravitation. It may
happen that the new state of nuclear matter (CGC) discovered in RHIC is
affected by the absence of borders or by the expansion of the universe. Even
though we do not have strong arguments against the cosmological QCD
transition, we will assume the hypothesis that it does not occur in the PNS
period.


\section{Cosmological consequences \label{sec-Conseq}}

\begin{figure}[ht]
\begin{center}
\includegraphics[height=6.4cm, width=8.5cm]{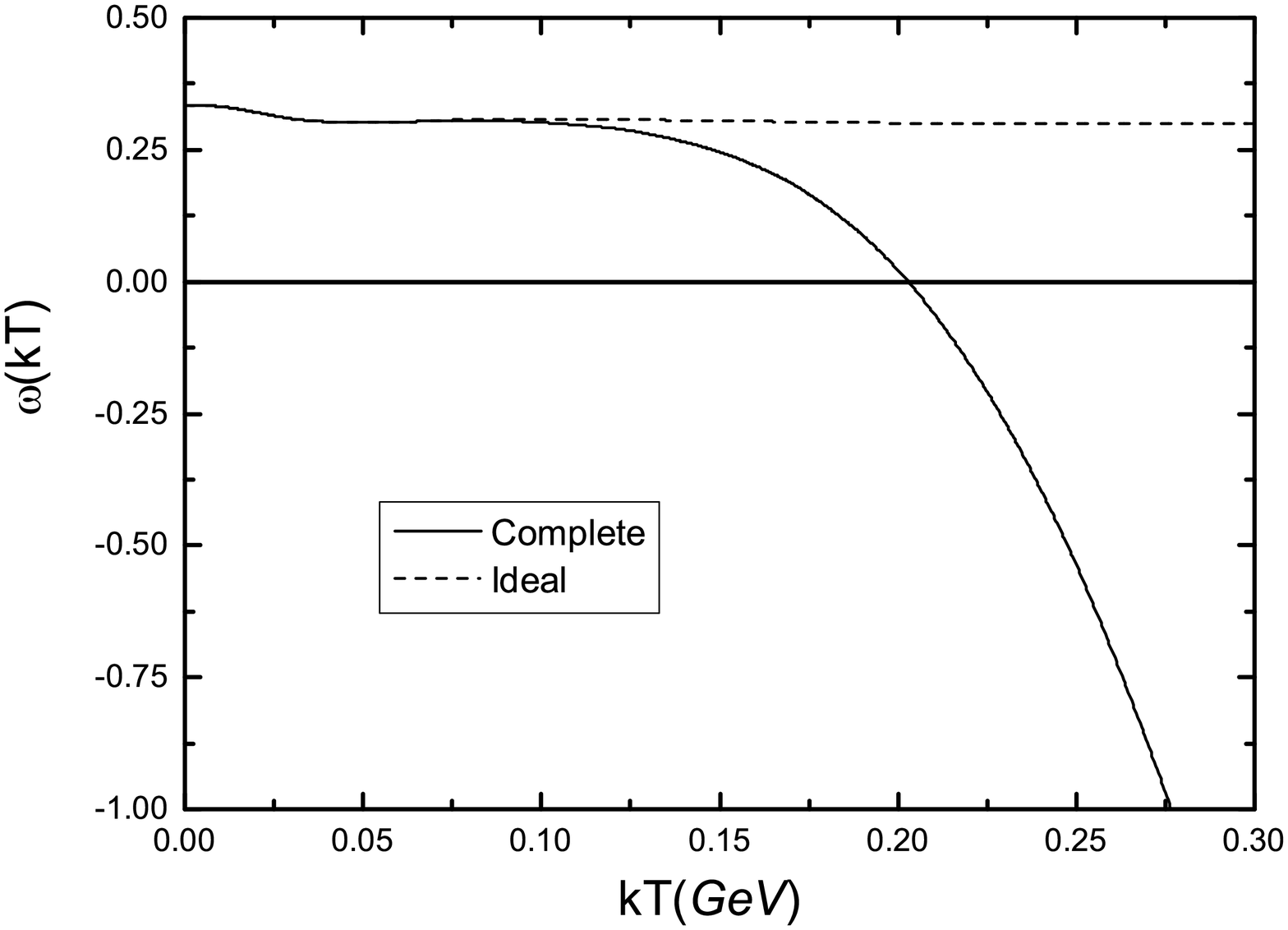}
\end{center}
\caption{Plot of $\protect\omega $ as a function of $kT$. Two curves are
shown: the complete $\protect\omega $ function, including all the particles
and relevant interactions (full line); ideal $\protect\omega $ function,
which computes all the particles but not the mutual interactions (dashed
line).}
\label{fig3}
\end{figure}

Let us admit that the proposed EOS, Eqs.~(\ref{C5_3}) and (\ref{C5_4}), are
valid during all the PNS period. A first application to cosmology can done
through the usual parametrization
\begin{equation}
p(kT)=\omega (kT)\rho (kT).  \label{C5_32}
\end{equation}%
Substituting (\ref{C5_32}) into the second Friedmann equation (\ref{C1_11}),
and neglecting $\Lambda $, we find
\begin{equation}
\left( \frac{\ddot{a}}{a}\right) = - \, \frac{4\pi G}{3}\left[ 1+3\omega (kT)%
\right] \rho (kT).  \label{C5_33}
\end{equation}%
As the energy density is a positive and increasing function of $kT$, the
information about the acceleration rate depends on the parametric function $%
\omega (kT)$\ solely. Isolating this quantity in (\ref{C5_32})\ and using
the explicit forms of $\rho (kT)$ and $p(kT)$, the plot of Figure \ref{fig3}
obtains. 
It shows that the complete and the ideal curves for $\omega $ are
practically superimposed until $0.11$ $GeV$. From this value on, the
complete $\omega $ curve begins to decrease, becoming negative at about $0.2$
$GeV$. On the other hand, the ideal $\omega $ curve remains positive and
nearly constant up to $0.3$ $GeV$. The effect of the behavior of the
complete $\omega $\ function on (\ref{C5_33}) is to change the universe's
expansion rate between $0.1$ $GeV<kT<0.3$ $GeV$. As we go back in time, the
universe passes continuously from a decelerated dynamics (pretty close to a
radiation dominated era) to an accelerated stage. Comparison of the two
lines in Figure \ref{fig3} shows that this acceleration is due to the
hadronic interaction processes.

The next step is to determine the scale factor $a(kT)$ and the function $%
t(kT)$\ associating cosmic time and temperature (or thermal energy). We
shall use the conservation equation,%
\begin{equation}
\frac{d\rho }{da}=-\frac{3}{a}(p+\rho ),  \label{Rtwenty three}
\end{equation}%
and the first Friedmann equation (\ref{C1_11}). We obtain $a(kT)$ by
integrating (\ref{Rtwenty three}):
\begin{equation}
a(kT)=a_{Pns0}\exp \left[ -\int\limits_{kT_{Pns0}}^{kT}\frac{\frac{d\rho }{%
d\left( kT^{\prime }\right) }}{3(p+\rho )}d\left( kT^{\prime }\right) \right]
,  \label{Rtwenty fiveA}
\end{equation}%
where the label $Pns0$ indicates that the quantity is calculated at the end
of the pre-nucleosynthesis period which, in energy, is $kT_{Pns0}=20MeV$.
The time function $t(kT)$\ is obtained by the combination of (\ref{Rtwenty
three}) and (\ref{C1_11}) ($c$ is re-inserted):
\begin{equation}
t_{Pns0}-t=\int\limits_{kT_{Pns0}}^{kT}\frac{\frac{d\rho }{d\left(
kT^{\prime }\right) }}{3(p+\rho )\sqrt{\frac{8\pi G}{3c^{2}}\rho -\frac{%
\kappa c^{2}}{a^{2}}}}d\left( kT^{\prime }\right) \text{,\ \ \ \ \ }\frac{%
8\pi G}{3c^{2}}=\frac{996\text{ }fm^{3}}{(ms)^{2}GeV}\,\,\,.
\label{Rtwenty six}
\end{equation}

Equations (\ref{Rtwenty fiveA}) and (\ref{Rtwenty six}) describe the
evolution of the pre-nucleosynthesis universe. A careful analysis of these
equations reveals that both diverge when $p= -\, \rho $\ and, according to
the complete EOS (with interactions included), this happens at $kT_{c}\simeq
0.2767$ $GeV$ (critical temperature).\footnote{%
Not to be confounded with the deconfinement temperature, also usually called
``critical temperature''.} The scale factor always decreases as $kT$\
increases. The decreasing rate $\frac{da}{d\left( kT\right) }$ is not so
well-behaved: it varies a lot and, as $kT$ approaches $kT_{c}$, $\frac{da}{%
d\left( kT\right) }\rightarrow -\infty $. This means that the \textit{scale
factor tends to zero when} $kT\rightarrow kT_{c}$. In order to study the
time function around $kT_{c}$ we have to particularize for the three values
of $\kappa $:\

\begin{enumerate}
\item For $\kappa =0$, the time interval $\left\vert
t_{Pns0}-t_{c}\right\vert \rightarrow \infty $ when $kT\rightarrow kT_{c}$,
because $\rho (kT) $ and $\frac{d\rho }{d\left( kT\right) }$ are always
positive. In a plane space-section universe, the time interval that would
elapse from the initial singularity till $t_{Pns0}$\ is infinity, and that
singularity is never attained.

\item For $\kappa =1$, we see from (\ref{C1_11})\ that it exists a value of $%
kT$\ close to $kT_{c}$\ where $\dot{a}=0$. This value, indicated by $kT_{B}$%
, is close to $kT_{c}$\ to $39$ decimal digits. Defining $a_{0}$\ as the
present--day value of the scale factor, and $a_{B}=a(kT_{B})$\ as the
minimum value of the scale factor, then it is found that $a_{B}\simeq
1.7\times 10^{-24}a_{0}$.\footnote{%
This calculation considers matter dominance in the interval $kT_{\gamma
0}\simeq 2\times 10^{-4}eV<kT_{\gamma }<$ $1$ $eV$ and radiation dominance
in $1$ $eV<kT_{\gamma }<$ $20$ $MeV$, with the initial condition $%
a(kT_{\gamma 0})=a_{0}$.} Thus, in an universe of spherical space-section,
the initial singularity does not exist; the universe reaches a minimum size
measured by $a_{B}$. Recall that recent cosmological data \cite{PDG}
slightly favor $\kappa =1$ (as $\Omega _{T0}\simeq 1.003$). Combining this
type of solution that presents a minimum size to the requirement that the
closed model presents also a maximum size, we obtain the so called \textit{%
eternal universe}, or \textit{bouncing universe \cite{Boucing}}. The time
interval from the minimum radius till the end of the PNS period is
determined by using (\ref{Rtwenty six}) and the values of $H_{0}$ and $%
\Omega _{T0}$; it is $\left\vert t_{Pns0}-t_{B}\right\vert \simeq 2.28$ $ms$.

\item For $\kappa =-1$, the time interval $\left\vert
t_{Pns0}-t_{c}\right\vert \rightarrow 2.28$ $ms$ when $kT\rightarrow kT_{c}$%
. Therefore, for a hyperbolic space-section, the initial singularity is
reached in a finite time.
\end{enumerate}

Notice that despite the different possible evolutions, the universe always
presents a \textit{maximum temperature} $kT_{c}$.

The statistical bootstrap model by Hagedorn also predicts a maximum
temperature $kT_{H}$ for the universe; but the mechanisms that lead to it
are distinct from the causes for our critical temperature $kT_{c}$. In the
context of Hagedorn's model, the increase in the total energy $E$\ is
responsible for the rise in the kinetic energy and also for the increase in
the number of kinds of particles in the system. As $E$\ rises it is more
advantageous to produce new particles than to increase the temperature of
the system. This ultimately results in an infinite limit for the energy ($%
E\rightarrow \infty $)\ and the pressure ($p\rightarrow \infty $) at a
finite value of the temperature ($kT_{H}<\infty $). On the other hand, $%
kT_{c}$\ is the temperature associated to the equality $p=-\,\rho $ in our
cosmological model with interacting particles -- see comments below Eq. (\ref%
{Rtwenty six}); $kT_{c}$\ appears due to the structure of the equations (\ref%
{C1_11}) and (\ref{Rtwenty three}).\


\subsection{The inflationary regime}


Function $\omega (kT)$ evolves continuously from $1/3$ (radiation era; $%
kT\leq 0.02$ $GeV$) and tends to $-1$ (as $kT\rightarrow kT_{c}$). The
inflationary regime (primeval accelerated expansion period) occurs when $%
\omega (kT)<-1/3$. On the other hand, our model requires that $\omega
(kT)>-1 $. We match these requirements by defining the inflationary period
as that for which $-1<\omega (kT)<-\frac{1}{3}$. According to Figure \ref%
{fig3}, this corresponds to the energy interval
\begin{equation}
0.2356\text{ }GeV\simeq kT_{accel}<kT<kT_{c}\simeq 0.2767GeV.  \label{C5_41}
\end{equation}%
An early acceleration is required to solve some cosmological problems --
horizon, flatness, origin of the inhomogeneities, etc. -- without imposing
specific initial conditions \cite{Guth,kolb,Liddle}. An accelerated regime,
however, must exhibit some special features to actually rule out these
problems. We now show that our model indeed eliminates the horizon and
flatness problems.

The \textit{horizon problem} is the lack of causal connection between
regions far apart which nevertheless exhibit similar physical
characteristics. The region of causal connection is quantified by the Hubble
radius $R\equiv (aH)^{-1}$, the maximal distance particles can travel during
an $e$-fold increase of the scale factor, an increase by the factor $e\simeq
2.78$ \cite{Dod}. The standard Bib Bang model tells us that, from the time $%
t_{Pns0}$\ until the present day, the maximum commoving distance between two
causal connected regions grew $9$ orders of magnitude:%
\begin{equation}
\frac{R_{0}}{R_{Pns0}}=\frac{a_{Pns0}H_{Pns0}}{a_{0}H_{0}}\sim 1.4\times
10^{9}.  \label{C5}
\end{equation}%
Let $kT_{hor}$\ be the energy value above which all the universe's content
is in causal contact. In order to solve the horizon problem in the context
of our model, we need to show that there is a $kT_{hor}<kT_{c}$ such that%
\begin{equation}
\frac{R_{Pns0}}{R_{hor}}=\frac{a\left( kT_{hor}\right) H(kT_{hor})}{%
a_{Pns0}H_{Pns0}}\sim 7.2\times 10^{-10}.  \label{C5_44}
\end{equation}%
This is done by using Eqs.(\ref{C1_11}) and (\ref{Rtwenty fiveA}). Indeed,
for all the values of $\kappa $, the $kT_{hor}$\ satisfying (\ref{C5_44})
respects $\left( kT_{c}-kT_{hor}\right) \simeq 10^{-35}$.

Analysis of the Hubble radius shows that, from the beginning of the
deceleration ($kT_{accel}\simeq 0.2356$ $GeV$) till nowadays, $R$ has
increased $1.75\times 10^{10}$. But from $kT_{hor}$ to $kT_{accel}$ the
Hubble radius diminished $1.8\times 10^{10}$. And for values of time smaller
than that corresponding to $kT_{hor}$\ all the universe was in causal
contact, allowing the thermalization of the cosmic fluid.

The \textit{flatness problem} may be treated quantitatively through the
first Friedmann equation%
\begin{equation}
1-\Omega _{T}=-\frac{\kappa c^{2}}{a^{2}H^{2}}=\Omega _{\kappa },\text{\ \ \
\ \ }\Omega _{T}\equiv \frac{\rho }{\rho _{c}}\equiv \frac{8\pi G\rho }{%
3c^{2}H^{2}}.  \label{C1_47}
\end{equation}%
According to recent data $\Omega _{T0}\simeq 1.003$. Now, $aH$ decreases
with time if the universe is of the radiation-dominated or matter-dominated
types (the usual cases). This means that $\left\vert \Omega _{\kappa
}\right\vert $\ increases with time in radiation and matter-dominated
models. If this is so, in the early universe $\left\vert \Omega _{\kappa
}\right\vert \ll $ $10^{-3}$ and we are led to the question: Why was the
total density value $\rho $ so close to the\ critical density $\rho _{c}$ in
the initial instants of the universe when $\kappa =\pm 1$? This necessary
fine-tuning in the initial condition for $\rho $ is the flatness problem.

The standard model establishes that the energy density at the end of the
pre-nucleosynthesis period $\rho _{Pns0}$ is close to $\rho _{c}$\ within a
precision of $21$ decimal digits. Hence, to solve the flatness problem one
needs to show that there is a value $kT_{flat}<kT_{c}$ such that%
\begin{equation}
\left\vert \Omega _{\kappa }\right\vert _{flat}=\frac{1.5\times
10^{-21}a_{Pns0}^{2}H_{Pns0}^{2}}{a^{2}(kT_{flat})H^{2}(kT_{flat})}\sim
10^{-3}.  \label{C5_42}
\end{equation}%
The value $kT_{flat}$\ verifying (\ref{C5_42}) is found to be $\left(
kT_{c}-kT_{flat}\right) \simeq 10^{-35}$\ (for $\kappa =\pm 1$). From $%
kT_{flat}$ to the end of the PNS, the scale factor increases $12$\ orders of
magnitude. And this large variation happens in a time interval of $2.25$ $ms$%
, by Eq. (\ref{Rtwenty six}). The duration of the accelerated expansion is
much smaller: $0.26$ $ms$. This makes clear that the high inflationary rate
-- $a(kT)$ increase of $10$ orders of magnitude in $0.26$ $ms$\ -- rapidly
flattens the universe.

The values (of the thermal energy, of the scale factor) characterizing the
PNS period in what concerns the inflationary issues are collected in Table 1.

\begin{equation*}
\begin{tabular}{|c|c|c|c|c|c|}
\hline
& $\Delta t(ms)$ & $\frac{a}{a_{Pns0}}$ & $\frac{R}{R_{Pns0}}$ & $\left\vert
\Omega _{\kappa }\right\vert $ & $k\left\vert T-T_{c}\right\vert (GeV)$ \\
\hline
$kT_{Pns0}$ & $0$ & $1$ & $1$ & $1.5\times 10^{-21}$ & $\simeq 0.2567$ \\
\hline
$kT_{accel}$ & $1.99$ & $5.0\times 10^{-2}$ & $8.0\times 10^{-2}$ & $%
9.55\times 10^{-24}$ & $\simeq 0.0411$ \\ \hline
$kT_{hor}$ & $2.25$ & $1.9\times 10^{-12}$ & $1.44\times 10^{9}$ & $10^{-3}$
& $10^{-35}$ \\ \hline
$kT_{flat}$ & $2.25$ & $1.9\times 10^{-12}$ & $1.44\times 10^{9}$ & $10^{-3}$
& $10^{-35}$ \\ \hline
\end{tabular}%
\end{equation*}

Table 1: Resum\'{e} of the main results concerning the behavior of our model
in the PNS period and the accelerated regime. $kT_{Pns0}=0.020$ $GeV$ is the
temperature at the end of the PNS; $kT_{acel}$ is the temperature below
which there is no acceleration; $kT_{hor}$ is the temperature at which the
causality problem is solved; and $kT_{flat}$ is the temperature that rules
out the fine-tuning in $\rho $. Definitions: $\Delta t=t_{Pns0}-t$ and $%
a_{Pns0}\equiv 1$.

\bigskip


\section{Final remarks \label{sec-Final}}


This work presents equations of state (EOS) for the early universe which
include the interactions among the constituent particles. We argue that the
dominant processes affecting the pre-nucleosyntesis period (PNS) are those
involving the strong interaction, the only which has been considered. Total
EOS accounting for photons, leptons and hadrons have been built, using a
phenomenological description of the hadronic interaction and assuming
thermodynamical equilibrium. Assuming the hypothesis of no deconfinement,
some interesting points have turned up.

First, the interactions naturally drove the pressure to negative values. The
de Sitter model has taught us long ago that an exotic equation of state, $%
p=-\rho $, in which the pressure is negative, would give rise to an
accelerated expanding universe. Therefore, our results could explain
the primeval acceleration regime, as an alternative scenario to
(scalar field) inflation. Indeed, our model is able to connect this
initial accelerated stage with a decelerated expansion of the type
expected for a radiation-dominated universe. In addition, the
acceleration generated in our interacting cosmic fluid has been
shown to be intense enough to solve the horizon and flatness
problems.

Another noticeable feature has been obtained: for the models with $\kappa =0$
or $\kappa =1$ the initial singularity is avoided, and a natural explanation
for the expansion of the universe emerges. In fact, either in an eternal
universe (compatible with $\kappa =1$) or in a universe with a beginning
(corresponding to $\kappa =0$), the primeval acceleration produced by the
strong interaction is capable of engendering an expansion which evolves to a
decelerated type and continue its dynamics. Nevertheless, let us emphasize
that these results have been derived for a symmetric universe: the quantity
of matter is assumed to equal that of antimatter.

Such effects are actually more general than the results of the specific
model proposed. Indeed, in order for then to be valid it is enough to
introduce in the FRW cosmology a continuous parameter $\lambda $\ (related
or not to interactions) responsible for the passage from $p(\lambda )=\frac{%
\rho (\lambda )}{3}$ to the exotic EOS\,\,\, $p(\lambda )=-\rho (\lambda )$
in such a way that $\frac{d\rho }{d\lambda }>0$ at all times. This type of
fitting links smoothly the accelerated and decelerated expansion periods via
the parametric equation $p=\omega \rho $ in which $\omega $ varies between $%
1/3$ and $-1$. This interval automatically eliminates the ghost models($%
\omega \leq -1$) \cite{ghost}.

Besides the mechanisms engendering primeval acceleration, there are
other important subjects related to the pre-nucleosynthesis period
to be discussed. Amongst them, we highlight two: the generation of
the initial perturbations of the energy density content and the
matter-antimatter asymmetry. The initial perturbations might be
generated\ in the context of our model through the introduction of
statistical fluctuation processes \cite{mag}; this possibility shall
be investigated in the future. The problem of the matter-antimatter
asymmetry is more involved: as it is well known, the inflation wipes
out all the traces of an eventual initial asymmetry and this demands
the existence of a mechanism driving a post-inflationary
matter-antimatter asymmetry. Moreover, a necessary condition to
produce an excess of particles over anti-particles is the system to
be out of the thermodynamical equilibrium \cite{Sak}. This exigence
prevents our model to describe this asymmetry once all the numerical
densities are obtained from the statistical mechanics in thermal
equilibrium. This limitation could be overcome if we consider
out-of-equilibrium phase-transition in a post-inflationary stage. A
mechanism of this kind was proposed in Ref. \cite{ucrafour} and will
the investigated in another opportunity.

From a wider perspective, this works calls attention to the fact that
particle interactions are a direct source of gravitation \cite{artigo 1}.
When one ignores this truth, simplifications result in the cosmological
models; but these are perhaps too large to enable a proper description of
the real universe. The role of the fundamental interactions in cosmology
opens new paths to the study of the universe evolution, and even if our
complete EOS are not valid during all the PNS period, the central idea may
be applied to model other eras. Up to this point, we would not dare to
affirm that the inclusion of hadronic interactions is more appropriated than
the scalar fields to generate the inflation. Many issues solved by the
inflationary theory have been left untouched here. But the interaction
scheme is certainly more suitable from the theoretical point of view: they
have a clear physical interpretation.


\section*{\protect\small Acknowledgements}

L. G. M. is grateful to Funda\c{c}\~{a}o de Amparo \`{a} Pesquisa do
Estado de S\~ao Paulo (FAPESP) and Funda\c{c}\~{a}o de Amparo \`{a}
Pesquisa do Estado do Rio de Janeiro (FAPERJ), Brazil; R. A. and R.
R. C. (grant 201375/2007-9) are thank to Conselho Nacional de
Pesquisas
(CNPq), Brazil. R. R. C. and L. G. M. also thank Instituto de F\'{\i}sica Te%
\'{o}rica, Universidade Estadual Paulista, Brazil, where this work was
initiated. Finally, R. R. C. would like to thank Prof. V. P. Frolov and Dr.
A. Zelnikov for the kind hospitality extended him at University of Alberta.

\appendix

\section{Multi-component systems \label{ap-Multicomp}}

A multi-component system is a set with more than one type of particle or
conserved quantum number. The grand canonical partition function $\Theta $
for a $B$-components system is, in analogy to (\ref{s1}),%
\begin{equation}
\Theta (z_{1},..,z_{B},V,T)=\sum\limits_{N_{1}=0}^{\infty
}...\sum\limits_{N_{B}=0}^{\infty
}Q_{N_{1},...,N_{B}}(V,T)z_{1}^{N_{1}}...z_{B}^{N_{B}}.  \label{D37}
\end{equation}%
(See Refs. \cite{Smith Lain 81,Osborn 77}.) The grand canonical potential $%
\Omega $\ as a function of the fugacities (and the temperature) is, then,%
\begin{equation}
\Omega (z_{1},...,z_{B},T)=\frac{1}{V}\ln \Theta
(z_{1},...,z_{B},V,T)=\sum\limits_{N_{1}=0}^{\infty
}...\sum\limits_{N_{B}=0}^{\infty
}b_{N_{1},...,N_{B}}z_{1}^{N_{1}}..z_{B}^{N_{B}},  \label{D39}
\end{equation}%
where $b_{N_{1},...,N_{B}}$ are the $B$-component cluster integrals, with $%
b_{0,...,0}\equiv 0$ ($N_{1}=N_{2}=...=N_{B}=0$).

The construction of the equations of state -- pressure $%
p=p(z_{1},..,z_{B},T) $, energy density $\rho =\rho (z_{1},..,z_{B},T)$ and
numerical densities $n_{1}(z_{1},..,z_{B},T)$, $n_{2}(z_{1},..,z_{B},T)$%
,..., $n_{B}(z_{1},..,z_{B},T)$ -- is done through the usual mapping from
statistical mechanics to thermodynamics, namely
\begin{subequations}
\begin{align}
\frac{p(z_{1},...,z_{B},T)}{kT}& \equiv \Omega
=\sum\limits_{N_{1}=0}^{\infty }...\sum\limits_{N_{B}=0}^{\infty
}b_{N_{1},...,N_{B}}z_{1}^{N_{1}}...z_{B}^{N_{B}},  \label{D41a} \\
n_{i}(z_{1},...,z_{B},T)& \equiv z_{i}\left( \frac{\partial \Omega }{%
\partial z_{i}}\right) _{V,T,\left\{ z_{B}\right\} \neq
z_{i}}=\sum\limits_{N_{1}=0}^{\infty }...\sum\limits_{N_{B}=0}^{\infty
}N_{i}b_{N_{1},...,N_{B}}z_{1}^{N_{1}}...z_{B}^{N_{B}},  \label{D41b} \\
\rho (z_{1},...,z_{B},T)& \equiv kT^{2}\left( \frac{\partial \Omega }{%
\partial T}\right) _{V,\left\{ z_{B}\right\} }=\left( kT\right)
^{2}\sum\limits_{N_{1}=0}^{\infty }...\sum\limits_{N_{B}=0}^{\infty }\frac{%
\partial b_{N_{1},...,N_{B}}}{\partial \left( kT\right) }%
z_{1}^{N_{1}}...z_{B}^{N_{B}}.  \label{D41c}
\end{align}%
Symbol $\left\{ z_{B}\right\} $ is the set of fugacities, from $z_{1}$ to $%
z_{B}$. As in the one-component case, these is another form for the EOS:
that corresponding to the virial expansion, with the pressure and the energy
density in terms of the numerical densities. It is:
\end{subequations}
\begin{subequations}
\begin{gather}
\frac{p(n_{1},...,n_{B},T)}{kT}=\sum\limits_{l_{1}=0}^{\infty
}...\sum\limits_{l_{B}=0}^{\infty
}a_{l_{1},...,l_{B}}(T)n_{1}^{l_{1}}...n_{B}^{l_{B}},  \label{D42a} \\
\frac{\rho (n_{1},...,n_{B},T)}{\left( kT\right) ^{2}}=\sum%
\limits_{l_{1}=0}^{\infty }...\sum\limits_{l_{B}=0}^{\infty
}c_{l_{1},...,l_{B}}(T)n_{1}^{l_{1}}...n_{B}^{l_{B}}.  \label{D42b}
\end{gather}%
The virial coefficients $a_{l_{1},...,l_{B}}$ and $c_{l_{1},...,l_{B}}$ are
determined by the clusters integrals $b_{N_{1},...,N_{B}}$. For a
2-component case, the first virial coefficients are
\end{subequations}
\begin{subequations}
\begin{align}
& \left. a_{0,0}=0,\text{\ \ \ \ \ \ \ }a_{0,1}=a_{1,0}=1,\right.
\label{ApenA1} \\
& \left. a_{2,0}=\frac{-b_{2,0}}{b_{1,0}^{2}},\text{\ \ }a_{0,2}=\frac{%
-b_{0,2}}{b_{0,1}^{2}},\text{\ \ }a_{1,1}=\frac{-b_{1,1}}{b_{1,0}b_{0,1}}%
,\right. \text{\ }  \label{ApenA2}
\end{align}%
\end{subequations}
\begin{subequations}
\begin{align}
& \left. c_{0,0}=0,\text{\ \ \ \ \ }c_{1,0}=\frac{\dot{b}_{1,0}}{b_{1,0}},%
\text{\ \ }c_{0,1}=\frac{\dot{b}_{0,1}}{b_{0,1}},\right.  \label{ApenA3} \\
& \left. c_{2,0}=-\dot{a}_{2,0},\text{\ \ }c_{0,2}=-\dot{a}_{0,2},\text{\ \ }%
c_{1,1}=-\dot{a}_{1,1}.\right.  \label{ApenA4}
\end{align}%
The dot $^{\cdot }$\ indicates differentiation with respect to $kT$.
Equation (\ref{Rnoventa dois A}) for a 2-component system reads:
\end{subequations}
\begin{subequations}
\begin{eqnarray}
b_{2,0}-b_{2,0}^{(0)} &=&\frac{1}{2\beta \pi ^{3}}\sum\limits_{a}\sum%
\limits_{l=0}^{\infty }g_{2,0}^{a}\left( 2l+1\right)
\int\limits_{2m_{X}}^{\infty }\omega ^{2}K_{2}(\beta \omega )\left[ \frac{%
\partial \delta _{l}^{a}(\omega )}{\partial \omega }\right] _{2,0}d\omega ,
\label{Aj8a} \\
b_{0,2}-b_{0,2}^{(0)} &=&\frac{1}{2\beta \pi ^{3}}\sum\limits_{a}\sum%
\limits_{l=0}^{\infty }g_{0,2}^{a}\left( 2l+1\right)
\int\limits_{2m_{Y}}^{\infty }\omega ^{2}K_{2}(\beta \omega )\left[ \frac{%
\partial \delta _{l}^{a}(\omega )}{\partial \omega }\right] _{0,2}d\omega ,
\label{Aj8b} \\
b_{1,1}-b_{1,1}^{(0)} &=&\frac{1}{2\beta \pi ^{3}}\sum\limits_{a}\sum%
\limits_{l=0}^{\infty }g_{1,1}^{a}\left( 2l+1\right)
\int\limits_{m_{X}+m_{Y}}^{\infty }\omega ^{2}K_{2}(\beta \omega )\left[
\frac{\partial \delta _{l}^{a}(\omega )}{\partial \omega }\right]
_{1,1}d\omega ,  \label{Aj8c}
\end{eqnarray}%
where $m_{X}$ and $m_{Y}$ are the masses of the first and second components.
The ideal terms are determined from the free-system \cite{Ruben1} as:
\end{subequations}
\begin{subequations}
\begin{align}
b_{1,0}& =\frac{g_{X}}{\Lambda _{X}^{3}(\beta )},\text{\ \ \ \ }b_{0,1}=%
\frac{g_{Y}}{\Lambda _{Y}^{3}(\beta )},\text{\ \ \ \ }b_{1,1}^{(0)}=0~;
\label{Aj11} \\
b_{2,0}^{(0)}& =\frac{(\pm 1)g_{X}}{2\Lambda _{X}^{3}(2\beta )},\text{\ \ }%
b_{0,2}^{(0)}=\frac{(\pm 1)g_{Y}}{2\Lambda _{Y}^{3}(2\beta )},  \label{Aj12}
\end{align}%
where $g_{X}$ and $g_{Y}$ are the number of internal degrees of freedom of
the free-systems; the upper (lower) signs refer to the Bose-Einstein
(Fermi-Dirac) statistics; and, $\Lambda (j\beta )$ is the relativistic
thermal wave-length\
\end{subequations}
\begin{equation}
\Lambda _{X,Y}^{3}(j\beta )=\frac{\left( 2\pi \right) ^{3}j\beta }{4\pi
m_{X,Y}^{2}K_{2}(j\beta m_{X,Y})}.  \label{Aj13}
\end{equation}%
The generalization for more than two components is straightforward.


\section{Hadronic scattering\label{ap-Espalha}}


Elastic hadronic scattering processes can be described by the phase shifts $%
\delta $ in the context of the partial--wave formalism. In this approach,
one uses the spectroscopic classification and separates the
\textquotedblleft elastic\textquotedblright\ process involving interactions
among pions, kaons and nucleons in six types.\footnote{%
The quotation marks were included to indicate that, besides the truly
elastic processes $\left( \pi ^{-}+p\rightarrow \pi ^{-}+p\right) $, we are
considering those with charge conjugation $\left( \pi ^{-}+p\rightarrow \pi
^{0}+n\right) $. The isospin symmetries $I$ are also taken into account.}
They are: $\pi \pi $, $\pi K$, $\pi N$, $KK$, $KN$ and $NN$. There is a set
of phase shifts $\delta $ for each of such processes. The $\delta $'s are
directly or indirectly determined from the experimental data. We will
discuss in detail how this is done in the case of the pion-pion scattering.

The \textquotedblleft elastic\textquotedblright\ pion-pion scattering
depends on the energy, the total orbital angular momentum $L$ and the total
isospin $I$.\ It is then convenient to introduce the notation $\delta _{L,I}$%
\ for each phase shift. As a scattering of identical particles with integral
spin, the associated two-pions total state (orbital angular momentum state
plus isospin state) must be symmetric. We will take $I=0$, $1$, $2$\ and
restrict the analyses to $L=0$, $1$, $2$ (partial waves $S$, $P$ and $D$).
This restriction is not arbitrary: it is imposed by the experimental data
that are available. Hence, the relevant phase shifts are those in Table B.1.

\begin{equation*}
\begin{tabular}{|c|c|c|}
\hline
$S\text{-Wave}$ & $P\text{-wave}$ & $D\text{-wave}$ \\ \hline
\begin{tabular}{c}
$\delta _{0,0}\text{\ \ with\ \ }g=1$ \\
$\delta _{0,2}\text{ \ with\ \ }g=5$%
\end{tabular}
& $\delta _{1,1}\text{ \ with\ \ }g=9$ &
\begin{tabular}{c}
$\delta _{2,0}\text{ \ with\ \ }g=5\text{ }$ \\
$\delta _{2,2}\text{ \ with\ \ }g=25$%
\end{tabular}
\\ \hline
\end{tabular}%
\end{equation*}
Table B.1: Phase shifts $\delta _{L,I}$ for $\pi \pi $-scattering. The
degeneracy degree $g$ accounts for the bosonic symmetry and the projections
of orbital angular momentum $\left( 2L+1\right) $ and of isospin $\left(
2I+1\right) $.

\bigskip

\begin{figure}[ht]
\begin{center}
\includegraphics[height=6cm, width=8cm]{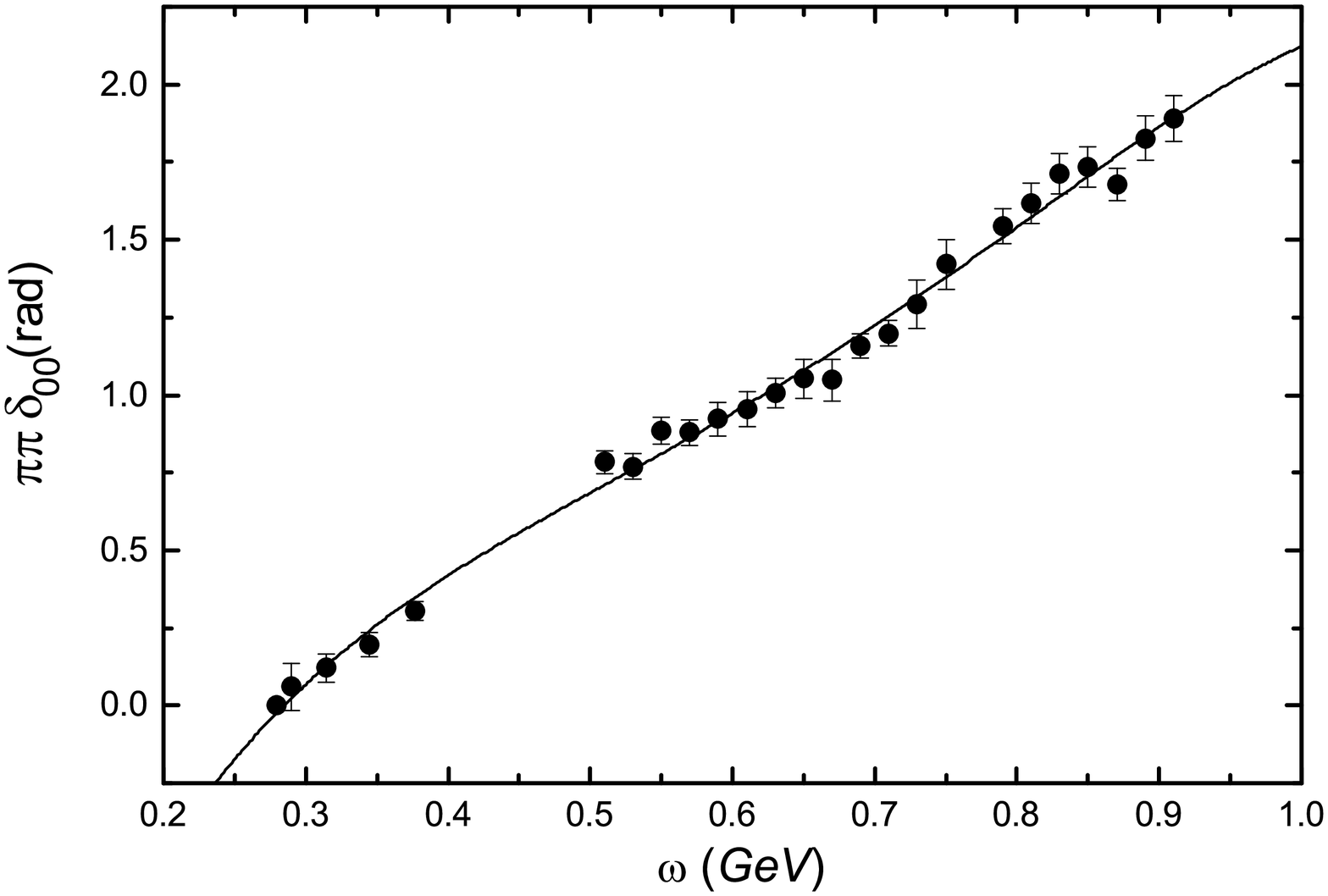} %
\includegraphics[height=6cm, width=8cm]{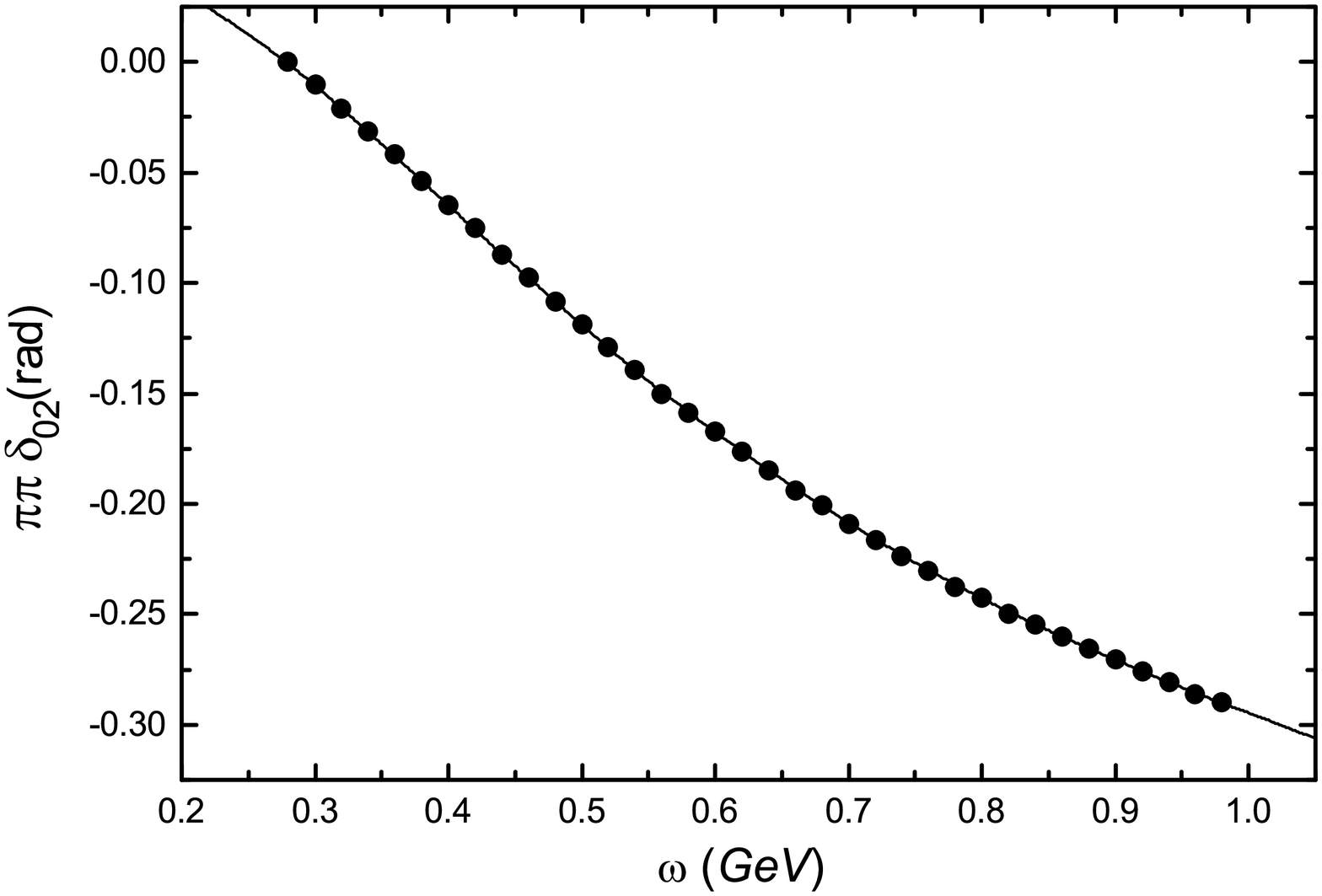}
\end{center}
\caption{Experimental data for the $S$-wave $\protect\pi \protect\pi $\
scattering and the corresponding fitted curves for the functions $\protect%
\delta _{L,I}\left( \protect\omega \right) $.}
\label{fig4}
\end{figure}

The data for the phase shifts $\delta _{0,0}$ and $\delta _{1,1}$ are found
in Ref. \cite{EstaMartin,Rosselet}, which give the results for the
extensively repeated and measured scattering $\pi ^{-}p\rightarrow \pi
^{-}\pi ^{+}n$. The data for $\delta _{0,2}$, $\delta _{2,0}$ and $\delta
_{2,2}$ \cite{EstaMartin,FrogPeter} are determined through modeling based on
the Roy equation \cite{Roy,Basdevant}. The $S$-wave data are shown in Figure %
\ref{fig4}.

The fits for each one of the phase shifts\ were done using the software
\textit{Origin }$6.1$ via the polynomial regression method or the non-linear
least squares fitting. The experimental uncertainties were considered. The
best-fit phase shifts are given by the expressions
\begin{equation}
\delta _{0,0}\left( \omega \right) =-1.13+4.36\omega -0.11\omega
^{2}-4.52\omega ^{3}+3.76\omega ^{4},  \label{ApF1}
\end{equation}%
and%
\begin{equation}
\delta _{0,2}\left( \omega \right) =-0.03+1.23\omega -6.67\omega
^{2}+12.62\omega ^{3}-13.39\omega ^{4}+8.57\omega ^{5}-3.11\omega
^{6}+0.48\omega ^{7}.  \label{ApF2}
\end{equation}%
Analogously, we can obtain the fitted curves $\delta _{L,I}=\delta
_{L,I}\left( \omega \right) $\ for partial waves $P$ and $D$. The functions $%
\delta _{L,I}\left( \omega \right) $\ -- such as (\ref{ApF1}) and (\ref{ApF2}%
) -- are differentiated with respect to the energy and substituted in
equations (\ref{Aj8a}-\ref{Aj8c}) for the cluster integrals $b$ which, in
turn, are used to obtain the $p$ and $\rho $ equations of state for the PNS
universe.

The other scatterings ($\pi K$, $\pi N$, $KK$, $KN$ and $NN$)\ are treated
in the same way. The experimental data used in the description of the
pion-kaon scattering are found in Ref. \cite{EstaCarnegie} and they are good
enough only to analyze the values $L=0$, $1$ -- $S$-waves and $P$-waves. In
the case of the pion-nucleon scattering, the relevant reference is \cite{CNS}
and we study $L=0$, $1$, $2$ ($S$, $P$ and $D$-waves). The kaon-kaon
scattering data are (indirectly) obtained from \cite{FrogPeter,KamiLesnLois}
and \cite{FurnLesn}\ using the separable potential formalism \cite%
{ActaPolonia,KamiLesn}; these data refer solely to the $S$-wave $K\bar{K}$
scattering (we suppose that the processes $K\bar{K}$ and $KK$\ are
identical, which means that the processes are independent of charge
conjugation $C$. The kaon-nucleon phase shifts (for $S$, $P$ and $D$-waves)
are in Ref. \cite{CNS} -- once again we admit independence under $C$. The
same Ref. \cite{CNS} presents the nucleon-nucleon data for $S$, $P$ and $D$%
-waves and, in these cases, in addition to the $C$-independence, it is
assumed independence on the isospin $I$: the proton-proton phase shifts are
very similar to the neutron-neutron ones.

\end{document}